\documentclass{aastex6}
\usepackage[T1]{fontenc}
\usepackage[latin9]{inputenc}
\usepackage{textcomp}
\usepackage{amsbsy}
\usepackage{amstext}
\usepackage{amssymb}
\usepackage{graphicx}

\makeatletter

\let\SF@@footnote\footnote
\def\footnote{\ifx\protect\@typeset@protect
    \expandafter\SF@@footnote
  \else
    \expandafter\SF@gobble@opt
  \fi
}
\expandafter\def\csname SF@gobble@opt \endcsname{\@ifnextchar[%]
  \SF@gobble@twobracket
  \@gobble
}
\edef\SF@gobble@opt{\noexpand\protect
  \expandafter\noexpand\csname SF@gobble@opt \endcsname}
\def\SF@gobble@twobracket[#1]#2{}

\makeatother

\usepackage[english]{babel}
\begin{document}

\title{Planetary System from the Outer Edge of the Inner Void\\
- Classes and Populations of Variety -}

\author{Masahiro Morikawa\footnote{{}hiro@phys.ocha.ac.jp} and Suzuka Amaya}

\affil{Department of Physics, Ochanomizu University, 2-1-1 Otsuka, Bunkyo,
Tokyo 112-8610, Japan}
\begin{abstract}
Planets are common objects in the Universe, observationally as well
as theoretically. However, the standard theory of their formation
encounters many difficulties, such as dust fall and disk lifetime
problems. We positively analyze them, expecting that those problems
as a whole may indicate some consistent effective model. Thus we propose
a dynamical model of the planet formation based on the assumption
that the inner void of gas is commonly formed in the disk without
specifying any Physical origin. The basic processes of this model
are the dust fall, the accumulation, and the slingshots. The dust
in the protoplanetary disk rapidly falls as it grows to the meter
size. Then, all of them stops at the outer edge of the void where
the gas friction disappears. Such dust clusters rapidly coalesce with
each other and easily cause the runaway in the dense and coherent
environment. Then the huge clusters are formed there and they are
the first generation planets Hot-Jupiters. They immediately slingshot
smaller clusters around them towards the outer regions. They are Rockey-Planets,
Cold-Gas-Giants, Ice-Giants, and `Trans Neptunian objects including
`Kuiper belt/Oort cloud objects`, depending on the original core mass
or the distance blown. Combining numerical calculations of the slingshots
and coagulation equations, we obtain the planet population diagram,
including the possibility of the massive thermal metamorphosis, the
origin of the variety of planetary systems, and the possibility of
stray planets/objects.
\end{abstract}

\section{Introduction - the origin of the variety -}

A variety of planets have been found by many advanced observational
methods \citet{Armitage(2019)}. We would like to know how are they
formed around a star, in particular the origin of their variety including
Hot-Jupiters, Cold-Gas-Giants, asteroids, comets, and stray interstellar
objects.

\textbf{} Planets are common structures in the Universe.\emph{ }The
recent discovery of thousands of planets \citet{NASA(2020)} itself
indicates that they are common. Moreover, from a simple theoretical
point of view, they are proved to be common. That is, the planets
are the stable structures of baryonic matter on the balance of their
quantum pressure and the self-gravity. We are already familiar with
another similar common structure in the Universe, the atoms. Atoms
are the stable structures of baryons and leptons on the balance of
their quantum pressure and the electromagnetism. Furthermore, the
atoms are naturally formed in the early stage of the Universe by the
Big Bang nucleosynthesis, except the tiny relative amount of heavier
atoms formed later in the various episodes and places. 

\textbf{}Therefore, we consider the planet formation process would
also be natural and common in the Universe. This is the first of our
guiding principle for constructing the planet formation models. Therefore,
we do not seek any elaborate mechanism nor detail of the individual
problems now but we try to extract any common indications from the
vast literature. Indeed, there have been many efforts to reveal the
origin of the planets for these several decades \citet{Armitage(2019)}. 

\textbf{}There seem to be continual difficulties that prevent the
natural formation of the planets. The essential difficulties are \citep{Ueda(2018)}:
Dust falls too fast toward the central star, in particular at the
meter size, before they grow to form protoplanets. Gas in the protoplanetary
disk disappears within $10^{7}$ years before all the giants are formed.
High-speed collisions and the inevitable turbulence in the gas destroy
the dust lump even once formed. Planets fall toward the center including
the type-II migration problem. Large collisions are needed for Earth
formation... and much more. All of these difficulties could be summarized
into the two facts: \emph{All object falls, time is short.} These
provide us with a good hint for constructing our scenario of planet
formation. If all dust falls, then no planets are formed at all. Therefore
there must be a void of gas at some region of the disk where no gas
friction works. This place should be very inside the disk to have
short time scales. Thus we propose a model assuming \emph{the existence
of a void of gas at the very central region of the protoplanetary
disk}. Then all the falling dust or lumps stop at the outer edge of
this inner void (OEIV) of gas. There may be some Physical mechanisms
that describe the formation of the gas void, such as Magnetorotational
instability, co-rotation instability, photoevaporation, etc. However,
we do not specify the mechanism in this paper to avoid complications
and the loss of generality. 

\begin{figure}
\includegraphics[width=17cm]{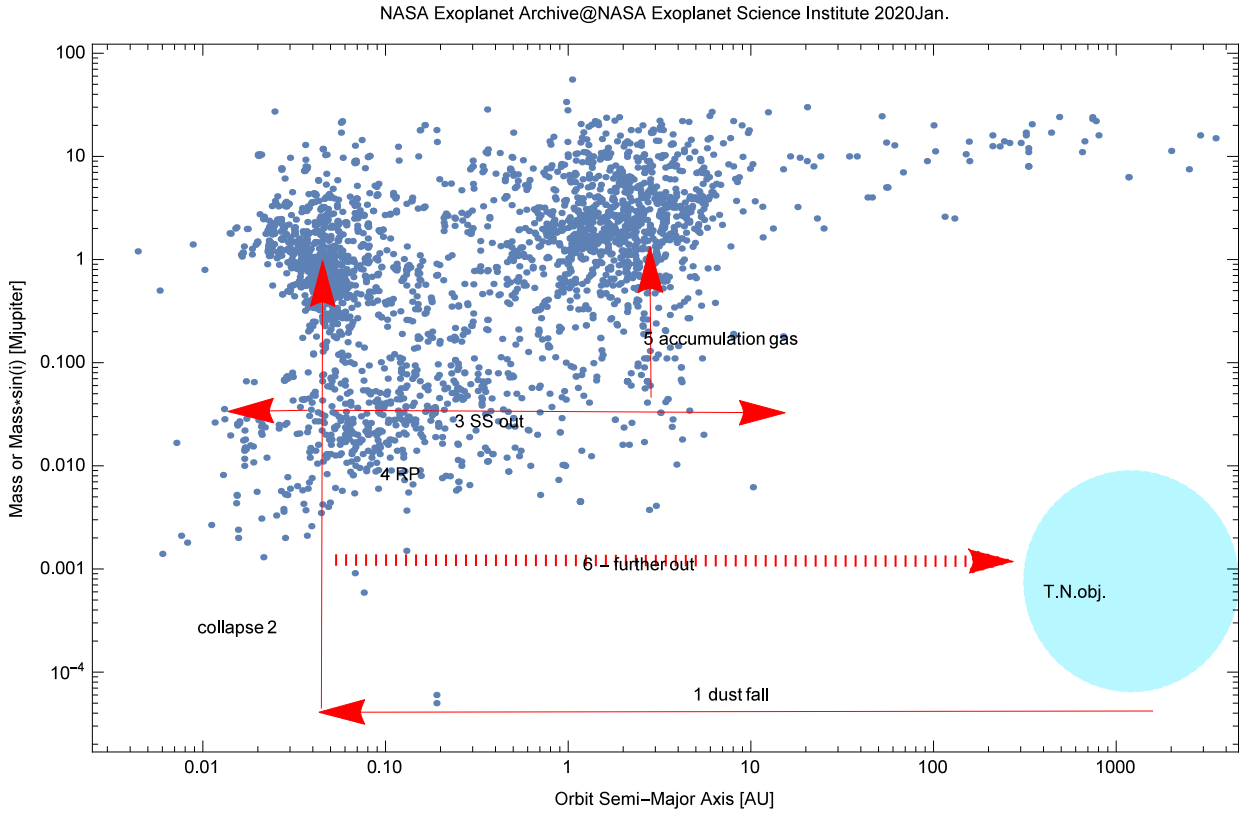} \caption{A variety of planets found so far is shown in the diagram of the mass
vs. semi-major axis\citep{NASA(2020)}. Our scenario of the planet
formation is illustrated on top of it with the label numbers corresponding
to the planet formation stages explained in the text. \label{fig1}}

\end{figure}

Thus, the stages of the planetary formation in our model goes as follows
in time sequence (see Fig.\ref{fig1}). The quantitative results are
explained in later sections. 
\begin{enumerate}
\item Dust and dust clusters\textbf{ fall toward the center but stops} at
the edge of the inner void and accumulate there. We analytically estimate
the size-dependent falling time scales. They may be very short such
as several hundred years. Turbulence and frequent destructive collision
with dissipation, on the common Kepler orbit, promote the coalescence
of those dust and dust clusters. 
\item The\textbf{ dense and coherent motion of dust/clusters at the edge
of the inner void induces the runaway growth} to yield several huge
planets. This population is the Hot-Jupiters. This orbit is too close,
say 0.04AU, to the center and Hot-Jupiters may not be stable. However,
this runaway time scale may be shorter and several hundred years based
on the standard disk model parameters\citep{Ueda(2018)}. \\
\item These massive planets at the outer edge \textbf{slingshot dust cluster
inward and outward}. This stage of evolution is essential to yielding
a variety of planet populations. 
\item Small dust clusters of size less than 10 Earth-mass are shot in and
out, but cannot effectively accrete gas, and the size of them cannot
grow. This population is the \textbf{Rockey-Planets}. They can be
shot widely to 0.01AU to 1000 AU. On the other hand, dust clusters
of size more than 10 Earth-mass can be shot up to 0.4AU by one Jupiter
(pure three-body system) and 1 to 100 AU by two Jupiters (pure four-body
system). These shot planets would migrate toward the center but shot
again outward; the planets are always dynamical objects regulated
by the outer edge of the inner void\citep{Ford(2001),Marzari(2002),Nagasawa(2008),Nagasawa(2011)}. 
\item \textbf{Large dust clusters of size more than 10 Earth-mass shut outside
can effectively accrete gas}, and the size of them can grow whenever
the gas exists there. In this process, their initially large eccentricity
dramatically reduces by acquiring the angular momentum of the gas.
This population is the Cold-Gas-Giants or Ice-Giants, depending on
the distance blown. They form gap and spiral structures on the planetary
disk. 
\item \textbf{Much smaller dust clusters can be shot very far and become
the seeds of `Trans-Neptunian Objects`}, including `Kuiper Best` of
exoplanet systems. The dust lump of $10^{12}$Kg, for example, can
be shot 100 to 1000AU within a time scale of 5000 years. Their orbital
inclination can easily exceed 30 degrees. Furthermore, the eccentricity
of them sometimes exceeds one. Therefore, a vast number of stray objects
in the interstellar space can be produced. 
\item Giants in term 4 would make gaps in the protoplanetary disk. In the
same logic as OEIV but with a much longer time scale, \textbf{outer
dust/clusters would fall toward the outer edge of these gaps and form
lumps}. These lumps may be scattered and resonated by Giants. Thus
we expect, in general, to have two kinds of small objects in a planetary
system, outer distributed lumps of hot origin and the inner distributed
lumps of cold origin.
\end{enumerate}
Although individual processes above would have already been studied
in the vast literature, sometimes in-depth, the whole scenario is
important to describe the variety of objects in a planetary system. 

The construction of this paper is as follows. In section 2, classes
and populations of planets, as well as of elements, are described
to show that the planets are common in the Universe. In section 3,
we describe the above stages in detail. In section 4, we discuss the
possible consequences of this scenario and some verifications of the
model. A variety of objects is also described here. Finally, in section
5, we summarize our work. 

\section{Classes and populations - planets are common - }

Planets are common objects in the Universe. Planets are balls of Hydrogen
or dust, in the first approximation. They are stable objects formed
by the balance of the gravitational attractive force and the quantum
mechanical repulsive effect. 

\subsection{Classes }

From the point of view of static stable structures, the planets, as
well as their constituents, would be categorized into the following
two classes. 

\subsubsection{Class I. Non-metal\protect\footnote{in astronomical sense.}}

Let us consider the hydrogen gas object first. The building block
is the Hydrogen atom whose structure is given by the equations, 

\begin{equation}
E_{atom}=\frac{p^{2}}{2m_{e}}-\frac{k_{0}e^{2}}{r},\quad\quad rp\approx\frac{\hbar}{2},
\end{equation}
where the first equation represents the electromagnetic energy $E_{atom}$
for an electron, $p$ and $r$ are, respectively, the electron momentum
and the position, $k_{0}$ is the strength of the electromagnetic
force, $m_{e}$ and $e$ are the electron mass and charge. The second
equation is the quantum uncertainly relation which is derived by the
$\mathrm{Schr\ddot{o}dinger}$ equation, by interpreting $p$ and
$r$ as the dispersion of the variables. $\hbar$ is the Planck constant
reduced. These equations determine the unique energy
\begin{equation}
E_{atom}=13.6eV,\label{eq:hatomenergy}
\end{equation}
corresponding to the bottom of the potential, as well as the mass
density of the object
\begin{equation}
\rho_{atom}=\frac{m_{p}}{\frac{4\pi}{3}r^{3}}=2.69\frac{gr}{cm^{3}}.\label{eq:hatomdensity}
\end{equation}
Using this density Eq.(\ref{eq:hatomdensity}), the balance of the
gravitational energy and the electromagnetic energy as a whole yields 

\begin{equation}
\frac{GM^{2}}{R}=E_{atom}\left(\frac{M}{m_{p}}\right)\label{eq:energy balance}
\end{equation}
where $G$ is the gravitational constant, $M,R$ are the object mass
and radius, $m_{p}$ is the proton mass, and therefore $M/m_{p}$
is the total number of Hydrogen. From this energy balance Eq.(\ref{eq:energy balance})
and mass radius relation 
\begin{equation}
M=\frac{4\pi}{3}R^{3}\rho_{\text{atom}},
\end{equation}
we have the scale of the object: 

\begin{equation}
R_{\text{planet }}=\frac{\hbar^{2}}{e\sqrt{2Gk_{0}}m_{e}m_{p}}=4.16\times10^{7}\mathrm{m}=0.59\mathrm{R}_{\text{Jupiter }},\label{eq:Jupiter R}
\end{equation}

\begin{equation}
M_{\text{planet }}=\frac{e^{3}k_{0}^{3/2}}{2\sqrt{2}G^{3/2}m_{p}^{2}}=8.12\times10^{26}\mathrm{Kg}=0.43M_{\text{Jupiter }}.\label{eq:Jupiter M}
\end{equation}
This is the structure of Jupiter. Thus, \emph{Jupiter is a common
object} in the Universe made from the balance of gravity and quantum
pressure. The thermal effect would certainly modify the above relation.
If this hydrogen ball can be considered as the ideal gas 
\begin{equation}
PV=Nk_{B}T,
\end{equation}
then this left-hand side, being the total energy, should be the same
as the left-hand side of Eq.(\ref{eq:energy balance}). Thus we may
conclude that the thermal effect becomes effective for $T\geq E_{atom}/k_{B}=1.58\times10^{5}K$
. Planets in our system are well below this temperature even at their
centers. 

\subsubsection{Class 2. Metals\protect\footnote{in the astronomical sense.}}

\begin{figure}
\includegraphics[width=17cm]{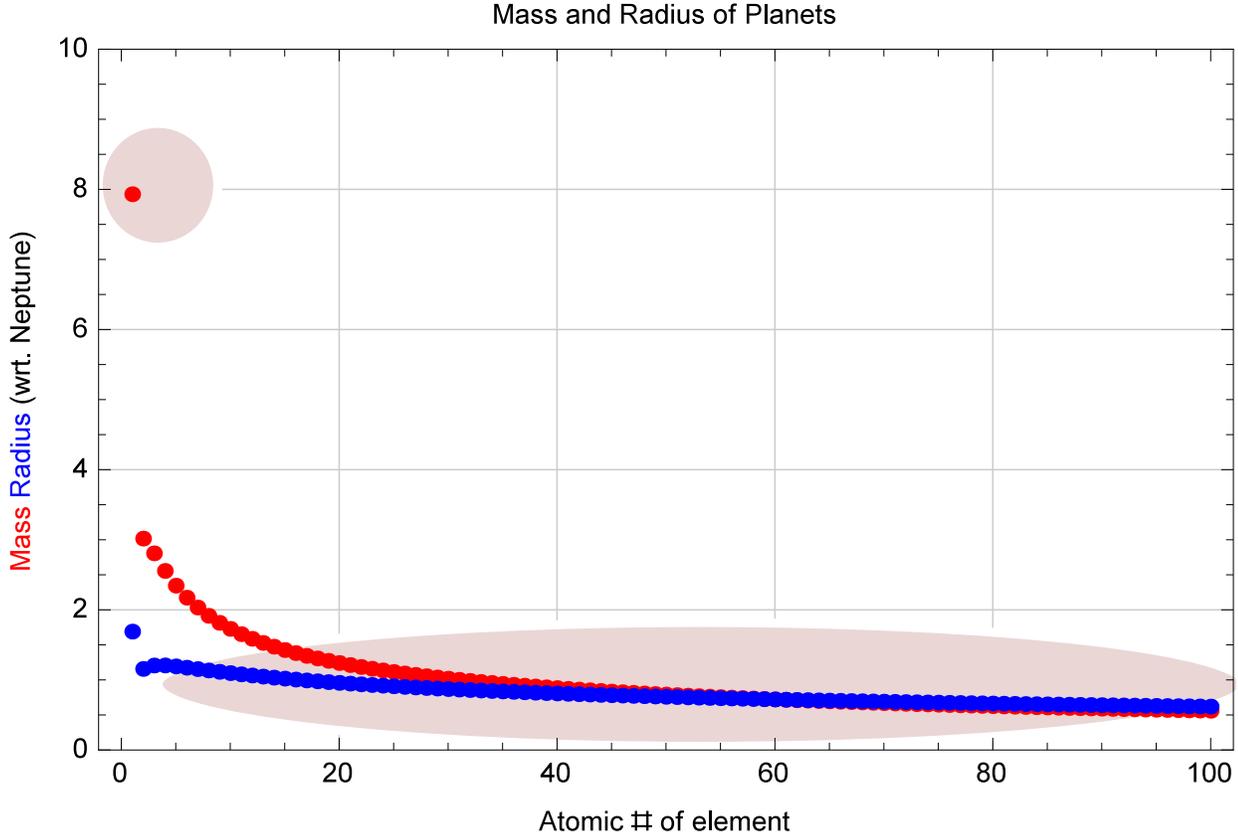}\caption{Mass(red) and radius(blue) of planets calculated from Eq.(\ref{eq:MRgeneral})
in the units of Neptune mass and radius. Two classes of planets are
clear. In the diagram, only the classes in mass is marked by pink
ellipsoids: Metal-ball class is almost Neptune and Non-metal-ball
class is almost Jupiter. \label{fig: massradius2}}
\end{figure}

If we further consider the object made from much heavier atoms of
atomic number $A$, the above argument is slightly modified. The energy,
with the uncertainty relation, 

\begin{equation}
E_{\text{atom }}=\frac{p^{2}}{2m_{e}}-\frac{k_{0}e^{2}A}{r},\quad\quad rp\approx\frac{\hbar}{2}
\end{equation}
should be extremized by the variation with respect to $r$. In the
present case, since we have many electrons, we add them one by one
on top of the previous ion, and sum over such extremized radius to
obtain the overall extension of the atom $r_{A}=\hbar^{2}H_{A}/e^{2}k_{0}m_{e}$,
where $H_{A}\equiv\sum_{n=1}^{A}n^{-1}$ is the A-th harmonic number.
This is a good approximation for the radius of each atom (\textit{cf.}
\citet{Cotton(1988)}) neglecting shell structures. Then the energy
is obtained as

\begin{equation}
E_{atom}[A]=\left(A-\frac{1}{2}\right)\frac{e^{4}k_{0}^{2}m_{e}}{\hbar^{2}H_{A}^{2}}.
\end{equation}
Approximating the atomic mass as $Z_{A}\approx2A-\delta(A-1)$, we
have the energy balace as before, 
\begin{equation}
\frac{GM^{2}}{R}=-E_{\text{atom }}\left(\frac{M}{Z_{A}m_{N}}\right),\frac{4\pi}{3}R^{3}\rho_{\text{atom }}.\label{eq:energy balance 2}
\end{equation}
where $\rho_{\text{atom}}$ is the atomic mass divided by the spherical
volume determined by the above $r_{A}$. From Eq.(\ref{eq:energy balance 2}),
we obtain the mass and radius of the object as

\begin{equation}
M=\frac{e^{3}k_{0}^{3/2}\left(2A-H_{A}^{-1}\right){}^{3/2}}{2\sqrt{2}G^{3/2}m_{p}^{2}(2A-\delta_{1-A}){}^{2}},R=\frac{\hbar^{2}\sqrt{H_{A}(2AH_{A}-1)}}{\sqrt{2Gk_{0}}em_{e}m_{p}(2A-\delta_{1-A})},\label{eq:MRgeneral}
\end{equation}
and shown in Fig.\ref{fig: massradius2}. 

As an example for $\mathrm{A}=26(\mathrm{Fe})$, we have the scale
of the object: 

\begin{equation}
\begin{array}{l}
R_{\mathrm{planet}}=0.90R_{Neptune},\\
M_{\mathrm{planet}}=1.09M_{Neptune},
\end{array}\label{eq:Neptune R and M}
\end{equation}
and even for $\mathrm{A}=113(\text{ Nihonium })$, 

\begin{equation}
\begin{array}{l}
R_{\text{phnet }}=0.61\mathrm{R}_{\text{Neptune}},\\
M_{\text{planet }}=0.53\mathrm{M}_{\text{Neptune}}.
\end{array}
\end{equation}
These are the structure of Neptune. Thus, \emph{Neptune is a common
scale} made from the balance of gravity and quantum pressure for heavier
atoms. These argument about the scales in the Universe is not exceptional
but all other cosmic structures can be obtained by the balance of
the gravity and the quantum pressure. 

\begin{figure}
\includegraphics[width=17cm]{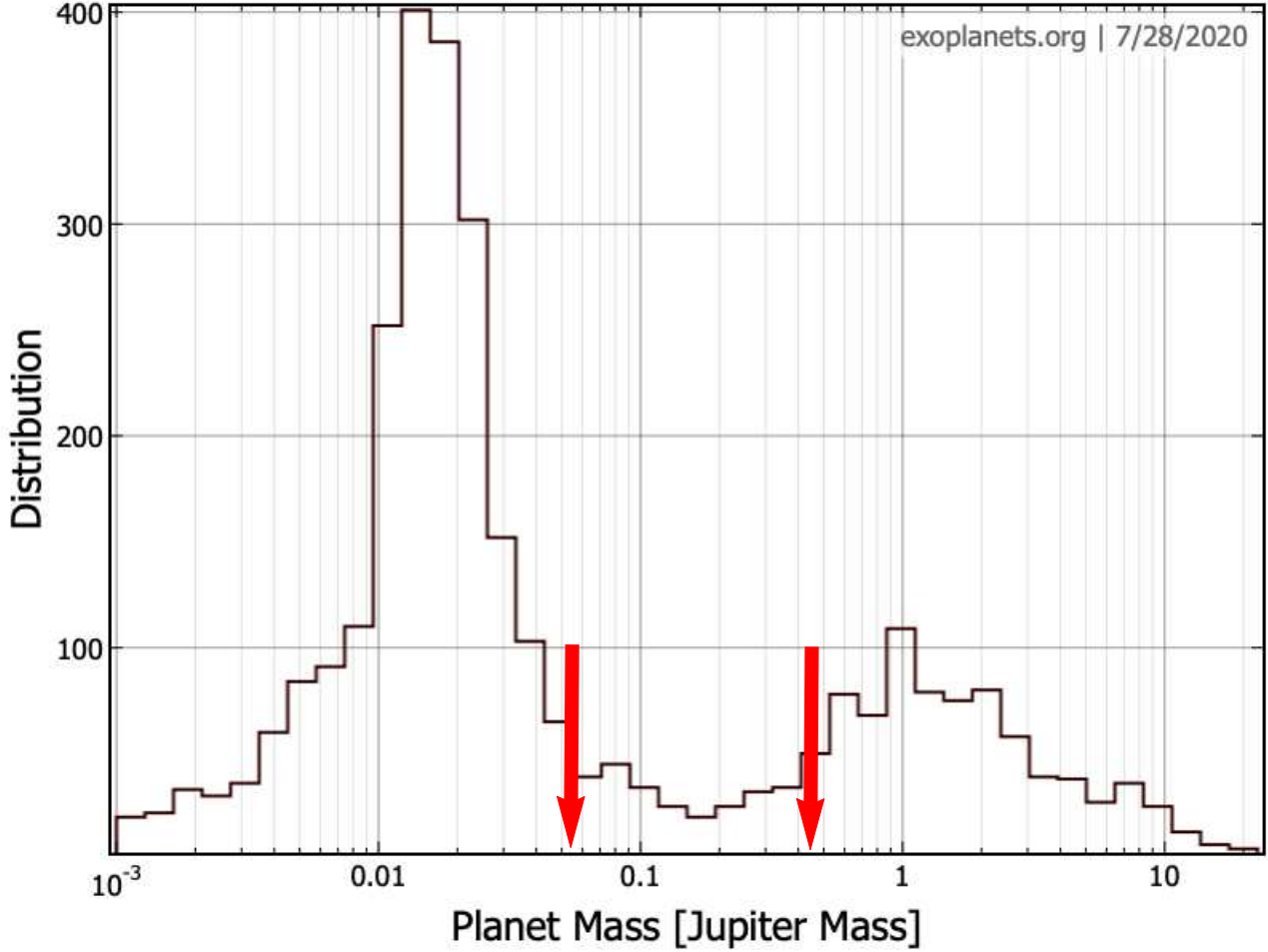}\caption{The mass function of the exoplanets observed so far\citep{NASA(2020)}.
The two arrows indicate the masses of the two classes: our 'Jupiter'
Eq.(\ref{eq:Jupiter M}) and 'Neptune' Eq.(\ref{eq:Neptune R and M}).
The data indicates the existence of two classes although our order
estimate deviates from their peaks by factors 2-3.\label{fig:The-mass-function}}
\end{figure}

In this way,\textbf{ Jupiter and Neptune are natural structures in
the Universe. They must commonly form without any elaborate mechanisms.
}The mass distribution of the observed exoplanets (Fig.\ref{fig:The-mass-function})
shows the double peak structure in accord with our argument. This
structure can also be interpreted as the structures of the core and
the core dressed by hydrogen gas. 

\subsection{Populations }

From the point of view of time evolution, the planets and their constituents
would be categorized into several populations. In the case of elements,
the populations are straightforward \citet{Nemiroff(2017)} and classified
as follows but they largely overlap with each other. 
\begin{description}
\item [{pop.I}] \textbf{Lightest elements H,He,(Li) (Big bang nucleosynthesis)}
They are formed within the first three minutes of the cosmic history. 
\item [{pop.II}] \textbf{C,N,O, F, Ne, Na,...Fe (Stellar nucleosynthesis)}
Pop-I elements cluster to form stars and nuclear fusion yields further
elements in its very inside. 
\item [{pop.III}] \textbf{..Co,Ni,...,Mo,... and much heavier elements
(Explosive nucleosynthesis).} A heavy enough star finally explodes
its envelope, leaving neutron stars and black holes, and much heavier
elements are formed. 
\item [{pop.IV}] \textbf{..Ru, Rh,...Pu (Merging neutron stars nucleosynthesis)
}These neutron stars collide with each other and yield further elements. 
\end{description}
We would like to categorize the planets as well. This is one of our
main objectives in this paper. We first have to know the history of
the planet formation. 

\section{Planetary formation stages }

We have shown a possible history of the planetary formation in the
protoplanetary disk. The history is made from the six stages: 1. (dustfall)
Dust and dust clusters fall toward the center but stops at the outer
edge of the inner void (OEIV) and accumulate there. 2. (rapid coagulation)
The dense and coherent motion of dust at OEIV induces the runaway
growth of several huge planets of size Jupiter. 3. (slingshots) These
massive planets slingshot dust clusters inward and outward. 4. (gas
accretion) Large dust clusters of size more than 10 Earth-mass shot
outside can efficiently accrete gas, and the size of them can grow.
5. (outer objects) Much smaller dust clusters can be shot very far
and become the seeds of `Trans-Neptunian Objects` including `Kuiper
Best` of exoplanet systems. 6. (secondary dustfall) Dust and dust
clusters at the outer region of the disk fall toward the center but
stops at the edge of the gap made at stage 4.

We will examine each of these stages in the following. 

\subsection{The edge of the inner void -falling dust and short time scale- }

Dust and dust clusters fall toward the center of the disk but stops
at the outer edge of the inner void (OEIV) and accumulate there. We
briefly estimate the size-dependent falling time scales. They may
be very short such as several hundred years of we set the radius of
the inner void as $0.04AU$. Turbulence and frequent destructive collisions,
with full dissipation on the common Kepler orbit, would train the
dust into coherent motion with small relative velocities and therefore
promote their efficient coalescence. 

\begin{figure}
\includegraphics{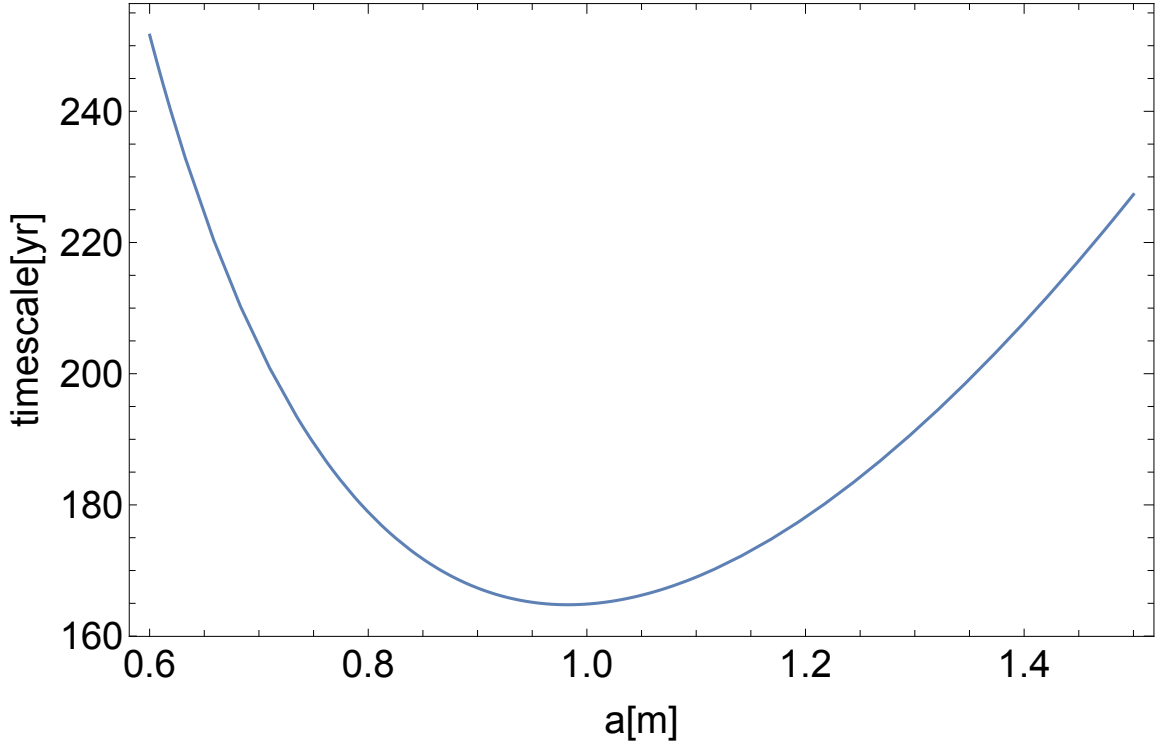}\caption{The dustfall time scale as a function of the radius of dust. The graph
is made by the numerical calculation of Eq.\ref{eq:dust dynamics},
with the parameter $\kappa=3.$ .\label{dust fall time scale} }
\end{figure}

A dust particle of mass $m$ with radius $a$ is described by the
Kepler equation with a drag force due to the gas (of dynamical viscosity
$\mu$),

\begin{equation}
m\frac{d^{2}\boldsymbol{r}}{dt^{2}}=-\frac{GM_{*}m}{r^{2}}\frac{\boldsymbol{r}}{r}-\kappa\left(\frac{d\boldsymbol{r}}{dt}-\boldsymbol{v}_{g}\right),\label{eq:dust dynamics}
\end{equation}
where $M_{*}$ is the mass of the central star, $\boldsymbol{v}_{g}$
is the gas velocity, $\kappa$ is the friction coefficient usually
expressed (the Stokes' law) as 
\begin{equation}
\kappa=6\pi\mu a,
\end{equation}
where $\mu$ is the viscosity and $a$ is the dust radius. We assume
the dust mass density $\rho_{0}=m/\left(4\pi a^{3}/3\right)$ is almos
constant. If the dust diameter $a$ is small and the friction force
dominates the inertial force, the two terms in RHS of Eq.(\ref{eq:dust dynamics})
balances with each other to yield a time scale 
\begin{equation}
\tau_{1}=\frac{\kappa\epsilon}{m\left(GM/r^{3}\right)}=\frac{9\mu\epsilon}{2\left(GM/r^{3}\right)\rho_{0}a^{2}}\propto a^{-2},
\end{equation}
where $\epsilon=\left(\frac{d\boldsymbol{r}}{dt}-\frac{d\boldsymbol{r}_{g}}{dt}\right)/\frac{d\boldsymbol{r}}{dt}$
and $\epsilon\approx\dot{10^{-3}}$. On the other hand, if the dust
diameter $a$ is large and the inertial force dominates, the balance
of the friction and the inertial term yields the time scale

\begin{equation}
\tau_{2}=\frac{m}{\kappa\epsilon}=\frac{2\rho_{0}a^{2}}{9\mu\epsilon}\propto a^{2}.
\end{equation}
Figure \ref{dust fall time scale} shows the whole behavior of the
dust falling time against the dust diameter $a$. These two time scales
are equal with each other $\tau_{1}=\tau_{2}$ at 
\begin{equation}
t_{ff}=\left(\frac{r^{3}}{GM}\right)^{1/2},\qquad a=\left(\frac{9\mu\epsilon}{2\rho\sqrt{GM/r^{3}}}\right)^{1/2}
\end{equation}
which roughly gives the shortest time scale of the dust fall and the
dust radius.  According to the detailed calculations\citep{Weidenschilling(1977)},
the meter scale dust cluster falls fasted with the time scale $10^{2}m/sec$.
Then this size of dust falls from the orbit of radius 1AU within 50
years. 
Thus, all the dust of any size falls toward the center. 

On the other hand, the planets are formed everywhere in the Universe
and therefore some portion of dust must remain escaping from the collision
to the central star. The only possibility for consistency is that
the falling dust should stop avoiding the gas friction somewhere around
the center of the disk. Therefore \textit{we assume the existence
of the void of gas at the center }of the protoplanetary disk. The
falling dust and clusters stop at the outer edge of the void and are
trained, by dissipative collisions and gravitational interactions,
to form coherent Kepler motion. 

There would be several reasons for the void to be formed: Magneto-Rotation
Instability, corotation instability, photoevaporation, etc. Each mechanism
of them would be quite complicated; the void may not be stable, may
evolve, and often disappears. We do not specify the mechanism of the
void formation but simply assume the existence of it in this paper. 

\subsection{Coherent Kepler motion to form Hot-Jupiters - runaway accretion - }

Dense and coherent Keplerian motion of dust at the outer edge of the
inner void (OEIV) induces the runaway growth of dust clusters as well
as the gravitational accretion of the gas. This process naturally
yields several huge planets of size Jupiter. This population is the
Hot-Jupiters, the first population of planets. This orbit is too close
to the center and Hot-Jupiters may not be stable due to the gravitational
interaction or the gas stripping off. However, the runaway time scale
and the slingshot time scale are shorter than the time scale of the
Hot-Jupiters, then the present scenario works and the Hot-Jupiter
population may not dominate the whole population at later stages. 

We now briefly estimate the accretion time scales. At the first stage,
the sticky dust growth would dominate before the gravitational accretion
dominates. The mass increase rate becomes 

\begin{equation}
\frac{dm}{dt}=\pi a^{2}\eta\rho_{\text{dust}}\Delta v.
\end{equation}
where $a=\left(\frac{3m}{4\pi\rho_{0}}\right)^{\frac{1}{3}}$is the
dust cluster radius, $\eta$ is the density henhancement factor due
to the dust fall to OEIV, and 
\begin{equation}
\Delta v\equiv-\frac{dv_{Kepler}}{dr}a=\frac{\sqrt{GM_{*}}a}{2r^{3/2}}
\end{equation}
is the typical relative velocity of the dusts. Then the time scale
of dust growth is 

\begin{equation}
\frac{m}{\dot{m}}=\frac{8\text{\ensuremath{\rho_{0}}}r^{3/2}}{3\eta\sqrt{GM_{*}}\rho_{\text{duat}}}.\label{eq:HJTimeradius-1}
\end{equation}
We use the standard disk model, 
\begin{equation}
\begin{array}{cc}
\rho_{gas}= & 2.0\times10^{-6}\left(\frac{r}{AU}\right)^{-11/4}\frac{\text{Kilogram}}{\text{m}^{3}},\\
\rho_{dust}= & 10^{-2}\rho_{gas},
\end{array}
\end{equation}
and setting $M_{*}$ as the solar mass, the mass density $\ensuremath{\rho_{0}=10^{3}kg/m^{3}}$
of the dust cluster itself, and $r=0.04AU$, the time scale of the
dust accretion becomes

\begin{equation}
\tau_{dustacc}=\frac{m}{\dot{m}}=243\left(\frac{100}{\eta}\right)\mathrm{year}.\label{eq:HJTimeradius}
\end{equation}

At the second stage, the mass increase rate by the gravitational gas
accretion becomes,

\begin{equation}
\frac{dm}{dt}=\pi r_{h}^{2}\rho_{gas}\Delta v,
\end{equation}
where $r_{h}=\left(\frac{r^{3}m}{3M_{*}}\right)^{\frac{1}{3}}$ is
the Hill radius. Setting $r=0.04AU$, we have the growing time scale
as 

\begin{equation}
\tau_{gasacc}=\frac{m}{\dot{m}}=\frac{6\sqrt{M_{*}/G}}{\pi r^{3/2}\rho_{gas}}=1.62\mathrm{year}.\label{eq:rHJTimehill}
\end{equation}
By a rough estimate, the micrometer size dust can form Jupiter within
$2.19\times10^{4}$ year, since the sticky dust growth is the bottleneck
of the process. If the radius of the falling cluster is much larger,
say $1$mm, the necessary time is $1.69\times10^{4}$ year.

Thus the Jupiter size objects, as well as any other scale objects,
can be easily formed in the OEIV. However, the actual estimate of
the time scale would be difficult since this OEIV region may be a
chaos by various gravitational interactions. Possible effects would
be as follows: 1) The dust density $\rho_{\text{duat}}$ should be
much higher since we have a continuous supply from falling dust. 2)
There may be a fair amount of dust that directly falls into the central
star. 3) Since dust is packed in a narrow region of OEIV, they scatter
with each other very often. This reduces the coherence of the common
Kepler motion and their orbits would be expanded. Therefore we would
eventually need N-body simulation for further detail. 

\subsection{\label{subsec:sling-shots-by}Slingshots by Hot-Jupiters - dynamical
origin of planets - }

The above first population planets, Hot-Jupiters formed at OEIV, yield
subsequent populations. These heavy planets slingshot other dust clusters
inward and outward. This stage of evolution is essential to yielding
a variety of planet populations. Small dust clusters of size less
than 10 Earth-mass are shot in and out, but cannot effectively accrete
gas and the size of them cannot grow. This population is the Rockey-Planets,
population II. They can be shot widely to 0.01AU to 1000 AU as we
will see soon. On the other hand, dust clusters of size more than
10 Earth-mass can be shot less effectively, but effectively acquire
gas and evolve to Cold-Giants, population III. These shot planets
would migrate toward the center but shot again outward at OEIV by
Hot-Jupiters if remain. Thus the planets are always dynamical objects
regulated by OEIV.

The slingshot dynamics are the many-body problem and the precise analytic
argument is impossible. Therefore we roughly estimate the basic feature
of the slingshot. Suppose we have two planets of mass $m_{p}$ which
slingshot the third object of mass $m$. We identify the maximum energy
extractable from the two planets as the transferred energy to the
third object. Then the maximum energy $E_{ss}$ used to the slingshot
will be the maximum potential energy released by the two planets of
mass $m_{p}$, and the escape energy $E_{cs}$ of the third object
is given by 

\begin{equation}
E_{ss}=\left(-\frac{Gm_{p}^{2}}{2r_{p}}\right)-\left(-\frac{Gm_{p}^{2}}{r_{h}}\right),\quad E_{cs}=-\frac{GM_{\odot}m}{r},\label{eq:SSenergy}
\end{equation}
 where $r_{p}$ is the radius of the planet and $r_{h}=\sqrt[3]{m_{p}r^{3}/(3M_{\odot})}$
is the Hill radius. If we choose the planet as Jupiter ($m_{p}$ is
the Jupiter mass), then we can estimate the maximum mass which can
be shot infinity as, 

\begin{equation}
m_{cr}=\frac{m_{p}^{2}r\left(r_{h}-2r_{p}\right)}{2M_{\odot}r_{h}r_{p}}=8.54M_{\oplus}\label{eq:SSmasscritical}
\end{equation}
where we used the radius of the OEIV as $r=0.04$AU. 

\begin{figure}
\includegraphics[width=17cm]{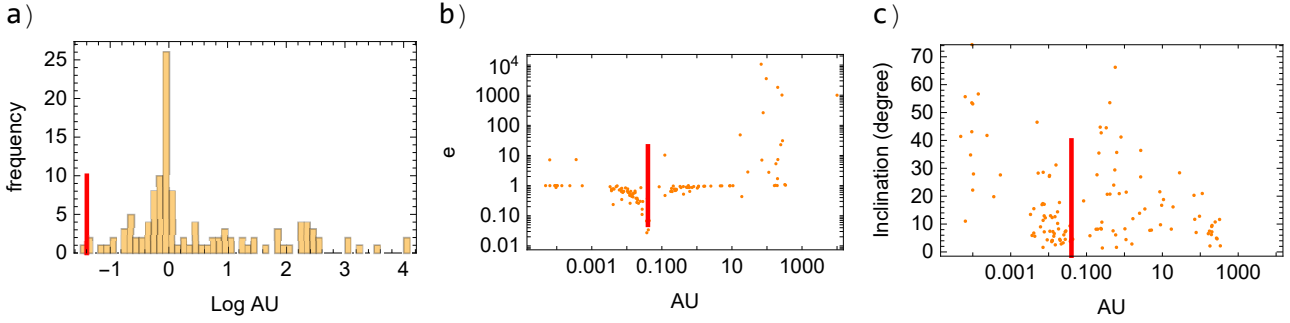}\caption{A lump of mass $10M_{\oplus}$ is shot by two Jupiters of mass $M_{J}$,
within 5150 years. a) frequency, b) eccentricity, and c) inclination
of the orbit. The location of the OEIV (0.04AU) is marked by a red
vertical bar. Most of the lump are shot to the distance 1 AU, but
about one-fourth shot to more than 100 AU. The initial locations of
all the bodies are randomly set within the three Hill radius of the
Jupiters, and the gravitational center of them is set at the distance
0.04AU from the central star. The initial velocities are set so that
they follow the circular motion. Another run with different separation
(several of Hill radius) doesn't show any significant difference in
the results.  We used the central star mass as the solar mass and
Jupiter mass density $10^{3}$Kg/m$^{3}$. Further, the force-cutoff
scale is simply set by the sum of the two radii of the approaching
bodies, assuming all have the same mass density. The numerical error
is well less than 1\%, evaluated from the energy conservation. \label{SS4-0.03-1-1}}
\end{figure}

However, the system is at least four body; the two planets, shot object,
and the Sun. The two planets can sink toward the central star and
can yield more energy to the slingshot. Actually, by the numerical
simulation as in Fig.(\ref{SS4-0.03-1-1}), this effect is strong
enough to slingshot $10M_{\oplus}$ object to $10^{3-4}$AU or more;
some portion of the run show the eccentricity of the shot object exceeds
one. 

The time scale of the slingshot would be very short for $r=0.04AU$;
the planets and the objects have a chance of slingshot per few days.
Therefore in the numerical calculations Fig.\ref{SS4-0.03-1-1}, we
set the whole period is 5150 years. 

If an object is shot to the radial distance $r$ from $r_{0}$, a
naturally expected eccentricity $\epsilon$ is, solving $\left(1+\epsilon\right)/\left(1-\epsilon\right)=r/r_{0}$,
given by $\epsilon=\left(r-r_{0}\right)/\left(r+r_{0}\right)$. This
tendency is manifest in the actual calculations in Fig.\ref{SS4-0.03-1-1}
and in the rest of the figures. 

\begin{figure}
\includegraphics[width=17cm]{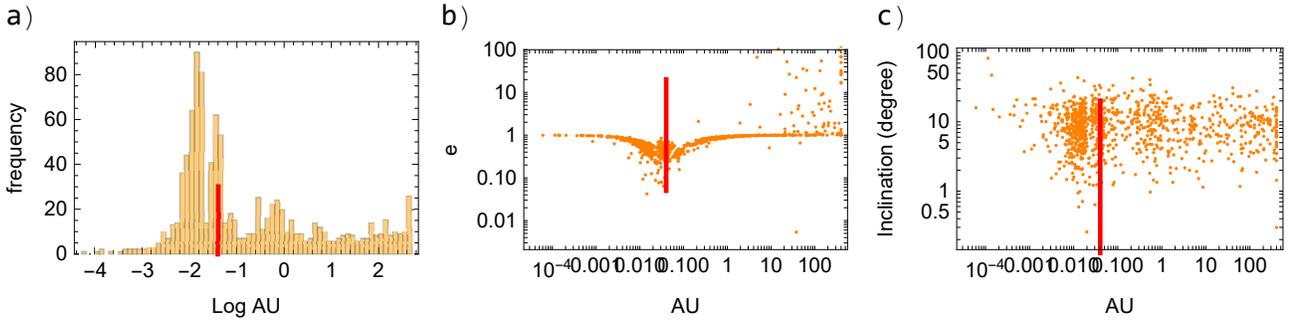}

\caption{Same as Fig.\ref{SS4-0.03-1-1}, but the lump of mass $10M_{\oplus}$
is shot by a single Jupiter of mass $M_{J}$. Most of the lump are
shot up to several handred AU. \label{SS3-0.03-1}}
\end{figure}

A slingshot of an object can be possible by a single planet. This
is a three-body problem: the central star, a planet, and the object.
The planet moves inward and yields energy transferred to the object.
We have tried numerical calculations and the results are shown in
Fig. \ref{SS3-0.03-1}. 

It turns out that \emph{the multiple objects in OEIV are unstable
}and cannot stay there forever. All the bodies, scattering and scattered
in OEIV are gravitationally unstable; slingshot is inevitable. This
instability is the essence of the present model to be common and universal.
Further, this instability implies the cases that the scattered objects
are commonly found but the scatterer can not be always found in OEIV;
it simply falls into the central star. 

\subsection{\label{subsec:rocky-planets--}Rocky-Planets - mass below $10M_{\oplus}$
- }

Relatively smaller dust clusters of size less than 10 Earth-mass shut
outside cannot effectively accrete gas, and the size of them cannot
grow \citet{Ida(2004)}. This is the population II planets. On the
other hand, the collision with gas would transfer the momentum and
the angular momentum from the gas to the planet. 

Gas pressure reduces the eccentricity of the rocky planets. Let us
estimate the time scale of the eccentricity reduction. We approximate
the planet orbit as Kepler and the radius is expresses as
\begin{equation}
r=\frac{L^{2}}{m^{2}GM_{*}}\left(1+\epsilon\cos\theta\right)^{-1},\label{eq:Keplerorbit}
\end{equation}
where $L$ is the orbital angular momentum, $\epsilon$ is the eccentricity:
\begin{equation}
\epsilon^{2}-1=\frac{2L^{2}E}{m^{3}G^{2}M_{*}^{2}},\label{eq:eccentricity}
\end{equation}
where $E$ is the total energy. Planet angular momentum $L$ and the
energy $E$ are perturbed by the surrounding gas in almost the Kepler
motion. Supposing the colliding gas simply transfer $L$ and $E$,
we have 

\begin{equation}
\begin{array}{cc}
\frac{dE}{dt}= & \frac{1}{2}\pi\rho_{gas}r_{p}^{2}\left(\Delta v\right)^{3},\\
\frac{dL}{dt}= & \pi\rho_{gas}r_{p}^{2}l\left(\Delta v\right)^{2},
\end{array}\label{eq:ELdot}
\end{equation}
where $r_{p}$ is the planet radius, $\Delta v$ is the planet velocity
relative to the ambient gas, and $l$ is the semi-major axes, given
by Eq.(\ref{eq:Keplerorbit}) setting $\theta=-\pi$. 

A time derivative of the both sides of Eq.(\ref{eq:eccentricity}),
using Eq.(\ref{eq:ELdot}), yields the time scale of the eccentricity
$\epsilon$ reduction as 
\begin{equation}
\tau=\frac{\epsilon}{-\dot{\epsilon}}=\frac{4G^{2}m^{3}M^{2}\epsilon^{2}}{\pi\text{\ensuremath{\Delta}v}^{2}\rho_{g}Lr_{pl}^{2}\left(-4El+\text{\ensuremath{\Delta}v}L\right)}.
\end{equation}
We express $\ensuremath{\Delta}v$ as the difference between the
slowest velocity of the planet at the end of the semi-major axis,
and the Kepler velocity there, which is expresses as 
\begin{equation}
\ensuremath{\Delta}v=\sqrt{\frac{GM}{l}}-\frac{L}{ml}.\label{eq:deltav}
\end{equation}

If we assume the late stage of the eccentricity reduction for the
Earth case (at $l=$1AU), we can apply the perturbation expansion
with respect to $\epsilon\ll1$. Then, $\ensuremath{\Delta}v=GmM_{*}\epsilon/(2L)$,
and 
\begin{equation}
\tau=\frac{4Gm^{3}M_{*}}{\pi r_{pl}^{2}\rho_{g}EL}.
\end{equation}
Using the gas density in the standard model,
\begin{equation}
\rho_{gas}=2.0\times10^{-6}\left(\frac{r}{AU}\right)^{-11/4}\frac{\text{Kilogram}}{\text{m}^{3}},
\end{equation}
setting $M_{*}$ as the solar mass, and using the Earth parameters,
we have the late stage eccentricity reduction time scale as $1.0\times10^{5}$
years.

If we assume the first stage of the eccentricity reduction for the
Earth case (at $l=$1AU), we can apply the perturbation expansion
with respect to $\left(1-\epsilon\right)\ll1$. Then, $\ensuremath{\Delta}v=GmM_{*}\sqrt{1-\epsilon(t)}/L$,
and 
\begin{equation}
\tau=\frac{4m\epsilon}{\pi r_{pl}^{2}\sqrt{GM/l}\rho_{g}\left(1-\epsilon-8\sqrt{1-\epsilon(t)}\right)}.\label{eq:time e-reduction deltav small}
\end{equation}
In the case of Earth, the initial eccentricity just after the slingshot
to 1AU from OEIV at 0.04AU, is expected to be given by the solution
of $\left(1+\epsilon\right)/\left(1-\epsilon\right)=1AU/0.04AU$ and
is $\epsilon=0.923$. Using this value to Eq.(\ref{eq:time e-reduction deltav small}),
we have the early stage eccentricity reduction time scale as $4.0\times10^{4}$
years.

All together considering the early and late stages, the eccentricity
reduction time scale for the Earth becomes $1.0\times10^{5}$ years.

\subsection{\label{subsec:gas-accretion-to}Gas accretion to form Cold-Giants
- mass over $10M_{\oplus}$- }

Large dust clusters of size more than 10 Earth-mass shut outside can
effectively accrete gas, and the size of them can efficiently grow
while the gas exists there \citet{Ida(2004)}. In this process, their
initially large eccentricity dramatically reduces by acquiring the
angular momentum of the gas. This population is the Cold-Gas-Giants
or Ice-Giants, depending on the distance blown. This is the population
III planets. They form gap and spiral structures in the planetary
disk. The mass increase rate by the gas accretion becomes

\begin{equation}
\frac{dm}{dt}=\pi r_{h}^{2}\rho_{gas}\Delta v,\label{eq:mdot}
\end{equation}
where the Hill radius $r_{h}$ is used because of the gravitational
accretion dominates now, and $\Delta v$ is the planet velocity relative
to the ambient gas.

If we estimate the time scale as before using Eq.(\ref{eq:deltav})
for $\Delta v$, we have
\begin{equation}
\tau=\frac{m}{\dot{m}}=\frac{3^{2/3}l^{1/2}M^{1/6}m^{1/3}}{\pi\sqrt{G}\rho_{g}r^{2}\left(1-\sqrt{1-\epsilon(t)}\right)}.
\end{equation}

If we assume the first stage of the mass accretion for the core of
10 Earth-mass cases at $l=r=$5.2AU, we have the mass increase time
scale as 

\begin{equation}
\tau=7.64\left(\frac{m}{10M_{earth}}\right)^{1/3}\text{yr}.
\end{equation}
The detail is much more complicated \citet{Kikuchi(2014),Higuchi(2017)}
and another tidal mechanism should also be considered for more accurate
calculations. Further, brown dwarves and Jupiter-planets are contaminated
with each other and could be separated by the eccentricity\citet{Bowler(2020)}
with the critical mass of about 15 Jupiter mass. 

Although it looks like we have circumvented the direct falling problem
in our model, the Type II migration process may force the mature Gas
Giants to fall toward the center. We would like to examine the possibility
to reconcile the problem. Gas Giants may fall, but at around the OEIV,
the secondary slingshot may happen to relocate this Giant to the outer
region. Thus shot Giants would show high eccentricity. 

\begin{figure}
\includegraphics[width=17cm]{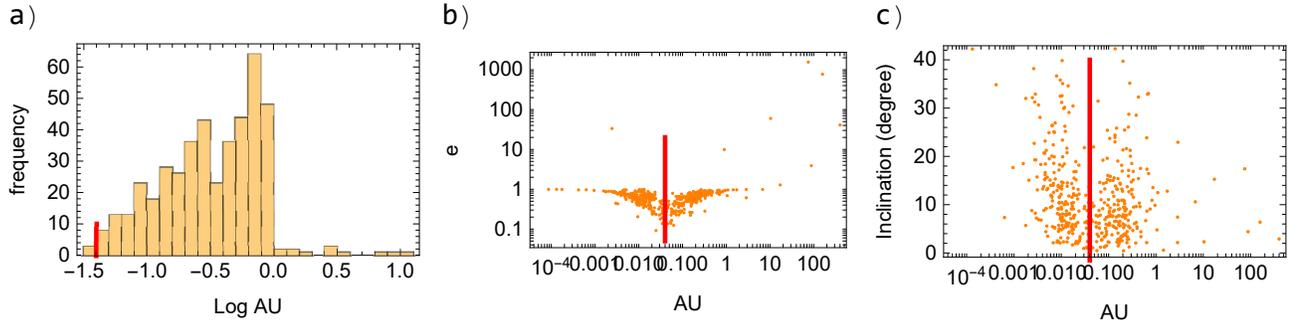}

\caption{A Jupiter is shot by the other two Jupiters. Most of the cases, the
Jupiter was shot up to 1 AU, although a small portion of the cases
shot to infinity. Therefore, the second slingshot for a Jupiter by
two Jupiters may not be effective. \label{3MJshot}}
\end{figure}

\begin{figure}

\includegraphics[width=17cm]{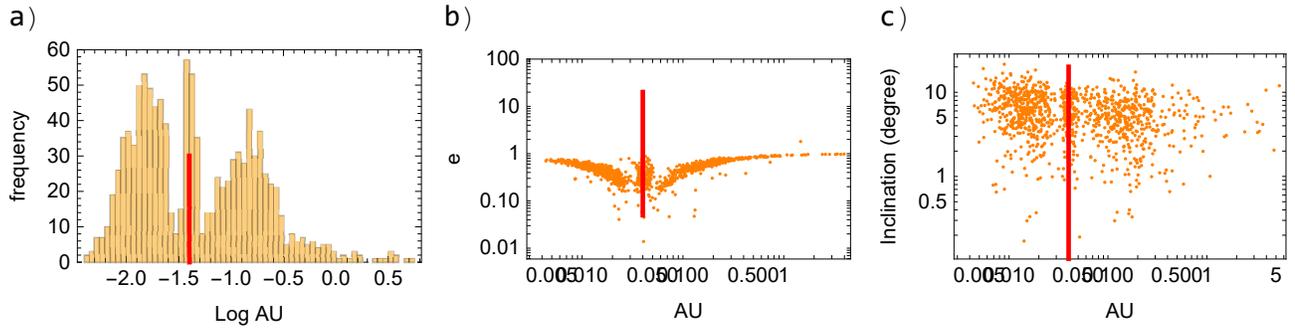}\caption{ A Jupiter is shot by another Jupiter. Most of the cases, the Jupiter
was shot up to 0.1 AU. Therefore, the second slingshot for a Jupiter
by another Jupiter is not effective. \label{2MJshot}}
\end{figure}

We have performed the similar numerical calculations of this case
in Figs.(\ref{3MJshot},\ref{2MJshot}). In most of the cases, Jupiter
was shot up to only 1 AU. Therefore, the secondary slingshot may not
properly work to maintain the Cold-Giants. On the other hand, the
secondary slingshot would be effective for much lighter rocky planets
provided the Hot-Jupiter was still survived. 

\subsection{\label{subsec:Strongly-shot-small}Strongly shot small objects -
Kuiper belt and Oort cloud objects - }

\begin{figure}

\includegraphics[width=17cm]{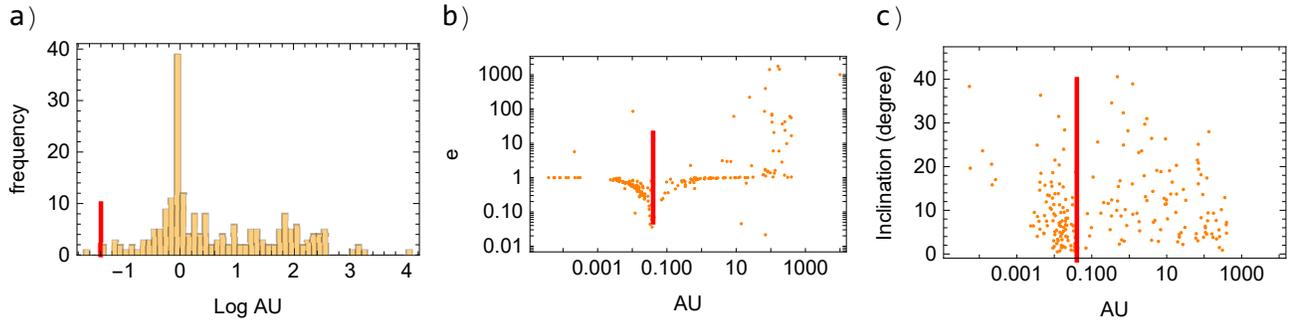}\caption{ Slingshot of the dust cluster of $10^{12}$Kg by two Hot-Jupiters.
It can be shot 100AU to 1000AU within a time scale of 5000 years.
Some cases of the run, the eccentricity of the dusct cluster well
exceeds one and easily escapes from the disk toward interstellar space.
\label{small cluster shot by 2 MJ}}
\end{figure}

\begin{figure}
\includegraphics[width=17cm]{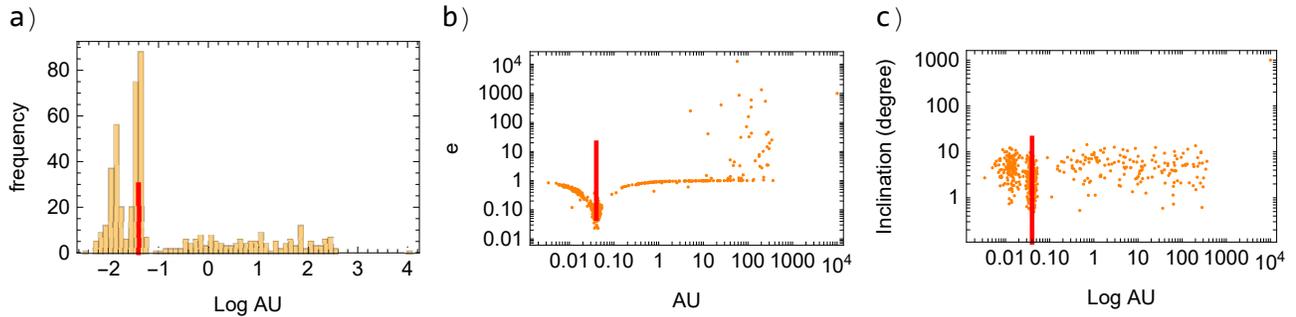}\caption{ Same as Fig.\ref{small cluster shot by 2 MJ} but the slingshot
by a single Hot-Jupiter. The dust cluster can be shot 1AU to several
100AU within a time scale of 5000 years. Some dust clusters can escape
from the disk toward interstellar space. \label{small cluster shot by 1 MJ}}

\end{figure}

Much smaller dust clusters can be shot farther out since the necessary
energy is smaller while the slingshot energy Eq.(\ref{eq:SSenergy})
is fixed. If they reach the edge of the disk, they may become the
seeds of `Trans-Neptunian Objects` or `Kuiper Best Objects` of the
exoplanet systems. Numerical calculations show that the dust cluster
of $10^{12}$Kg, for example, can be shot 100 to 1000AU by two Hot-Jupiters
within a time scale of 5000 years as shown in Fig.(\ref{small cluster shot by 2 MJ}).
Their orbital inclination can easily exceed 30 degrees. 
Even a single HJ can shoot the dust cluster to the edge of the disk
as shown in Fig.(\ref{small cluster shot by 1 MJ}). 
Based on the above mechanism, we now introduce the two new generations:
the object shot inside the disk (population IV), and the object shot
outside (population V).

Small clusters that remain inside the disk near the outer boundary
fully attract ice-water by taking a long time. These belong to the
population IV in the history of planet formation. Since they are directly
shot from the central region of the disk and therefore they inevitably
experienced the very hot environments of temperature about 2000K.
Thus, they all contain the thermally metamorphosed components at least
in their cores. 

Small clusters that escape from the disk stray in the interplanetary
space. Actually in our numerical calculations Figs.\ref{small cluster shot by 2 MJ}-\ref{small cluster shot by 1 MJ},
the eccentricity of a fair amount of them exceeds one in both cases
of the shot by two and a single Hot-Jupiters. Therefore, a vast number
of stray objects in the interstellar space are expected. Also from
the fact that the mass of these populations well below the critical
mass Eq.(\ref{eq:SSmasscritical}), and therefore plenty of stray
small planets are expected. They are relatively native objects directly
shot from the center of the disk without strong contamination from
the environments. Therefore, the whole body of them will be thermally
metamorphosed. 

\subsection{\label{subsec:Fall-toward-the}Fall toward the outer edge of the
gap - main-belt objects - }

A Cold-Giant in \ref{subsec:gas-accretion-to} would make a gap in
the protoplanetary disk. In the same logic as the case of OEIV but
with a much larger scale, outer dust may fall toward the outer edge
of this disk gap (OEDG) and accumulate on this orbit. They form small
dust lumps there. If this place is outside of the snow-line, water-ice
would accrete on these lumps. This is the population-VI object. These
small lumps may be scattered and resonated by Giants. 

Thus we expect to have two kinds of small objects in a planetary system,
a) outer distributed clusters of hot origin \ref{subsec:Strongly-shot-small},
and b) the inner distributed clusters of cold origin. These two kinds
of small objects are easily disturbed or resonated by other planets
and eventually settle to some stable places. In these places, the
above two kinds of small objects co-exist, collide, and often coalesce
with each other. 

\section{Discussions}

We may be able to check the validity of the present model from the
following considerations although the precise validation would be
premature for our rough model.

\subsection{\label{subsec:Populations-of-planets}Populations of planets }

We now summarize the populations of planets base on our model of planet
formation. For each population, we try to extract any useful information
for the verification of our model. 

\textbf{Population I. Hot-Jupiters(HJ) }The first formed planets are
the Hot-Jupiters (HJ) at the outer edge of the inner void(OEIV). Dust
and the dust clusters fall toward OEIV and accumulate there. They
collapse into the protoplanets in various sizes. In this condensed
coherent orbit on OEIV, runaway collapse easily takes place to form
a huge planet of size Jupiter and flick off the other smaller objects
inward and outward by slingshot. At the same time, the orbit is unstable
against mechanical interactions, and the HJ themselves can be shot
inner or outer. Therefore HJ, despite the first formed planet, may
not stay on the original orbit. The time scale of HJ formation, firstly
by the dust accumulation and secondly by gravity, is about several
hundred years. The actual time scale of HJ formation from micron-scale
dust is about $2.2\times10^{4}$ years. This rapid formation is possible
because we set the radius of OEIV as 0.04AU and the dust accumulation
effect. 

\textbf{Population II. Rocky-Planets(RP) }Dust cluster of mass less
than 10 Earth-mass shot by HJ from the OEIV would form Rocky-Planets
(RP). The earth belongs to this population. They cannot get gas accretion,
but reduce their orbital eccentricity, which was initially large,
by the interaction with the gas. A typical time scale of the eccentricity
reduction is about $10^{5}$years. 

On the other hand, recent observation of the exoplanets shows a fair
amount of RP have eccentricity more than 0.2\citet{NASA(2020)}. This
fact appears to conflict with the above small time scale of eccentricity
reduction. However, the slingshot at UEIV would repeat many times,
and the final slingshot just before the gas depletion would finally
set the persistent eccentricity, which we now observe. 

\textbf{Population III. Cold-Giants(CG)} Dust cluster of mass larger
than 10 Earth-mass shot by HJ from the OEIV would form the core of
Cold-Giants (CG). The cores can accrete gas to grow to form the full
body of CG. Jupiter belongs to this population. The time scale of
the CG formation by gravity is very short about ten years. The eccentricity
reduction time scale should be the same order.

On the other hand, a fair amount of CG has eccentricity more than
0.4 as well \citet{NASA(2020)}. Unfortunately, we cannot reconcile
this fact with the above short time scale by the repeated slingshots
as in the case for the population-II. This is because the shot distance
for CG is limited up to about 1 AU. (see. Figs.\ref{3MJshot}-\ref{2MJshot}) 

\textbf{Population IV.} \textbf{`Trans-Neptunian` and `Farthest` objects
(TN) }Lighter objects can be strongly shot to the outer region in
the disk. Kuiper belt and Oort cloud objects belong to this population,
although most of them would be a mixture with the population VI objects
explained next.

\textbf{Population V. Stray objects }Smaller the object, more chance
to be shot distant and some may escape the disk to form stray objects.
1I/2017 U1or Oumuamua is the first detected such interstellar object
\citet{Meech(2017)}. 2I/Borisov or Borisov comet is the first detected
such interstellar comet \citet{King(2019)}. More objects of these
kinds \citet{Namouni(2020)}would be expected from our simple numerical
calculations. 

\textbf{Population VI.} \textbf{asteroids/comets} CG in the gas disk
generally forms gap. Dust and clusters located outside of the gap
would fall toward the outer edge of this disk gap (OEDG) formed by
CG and stop there. The situation is the same as OEIV but with a longer
time scale and smaller mass scale due to the dilute dust density and
the weak gravity. The Main Belt objects in our solar system belong
to this population, possibly being the mixture with pop.IV. 

Despite the above classifications of the planetary populations, they
would be mixed and contaminated with each other. A typical example
of this mixture would be the populations-IV and VI. Small objects
are generally the contamination of the hot origin (pop IV) and the
cold origin (pop VI). These objects further contaminate the population
II planets. In this way, the variety of the planetary system expands
beyond the original populations. 

Although the existence of the above variety, we expect that the cores
of the planets, excluding the accumulated component later, would have
common composition. This is because all the cores are formed at a
single place the OEIV with strong mixing in our model. 

Recent observation \citet{Doyle(2019)}of the six white dwarf atmosphere
spectrum reveals that the chemical composition and the Oxygen fugacities
relative to IW ($\Delta IW$) are almost the same as those of our
Solar system including Earth, Mars, Mercury, and typical asteroids
in the Solar System. The white dwarf atmosphere is expected to contain
the destroyed and evaporated planet cores after the late-stage expansion
activity of the central star which finally formed the white dwarf.
Thus the rocky exoplanets, so far observed, are geophysically and
geochemically similar to the Solar System, suggesting the planetary
systems are common. 

\subsection{\label{subsec:Have-all-cores}Have all cores and small objects undergone
thermal metamorphism?}

The cores of the planets, in our model, should initially have experienced
the thermal metamorphosis. This is because all the planet cores are
formed at the outer edge of the inner void (OEIV) very near the central
star. The equilibrium temperature becomes 1400K for the planet at
0.04AU around the Sun, estimated from the energy balance equation
$T_{eq}=\sqrt{R_{*}/(2r)}T_{*}$, where$T_{*},R_{*},r$ are the surface
temperature and radius of the central star and the orbital radius
of the planet, respectively. The temperature would further increase
by the frequent mutual collisions at OEIV before the completion of
planet formation. After that, the core is slingshot outward and cools
monotonically. This thermal metamorphism history applies to most of
the populations except population VI (clustering at OEDG); applies
to `Trans-Neptunian objects` of population IV, and to some `Asteroid
belt objects` which are considered to be the mixture of populations
IV and VI. 

The study on comets, meteorites, and asteroids of our solar system
may provide some relevant information for this scenario.

In the case of \textit{comets}, plenty of water often found in them
cannot be maintained at the place only 0.04 AU separated from the
central star, suggesting that the population IV is naturally excluded
as their origin. However, the Stardust Mission \citet{Brownlee(2014)}
has revealed that the dust of comet Wild 2 includes the high-temperature
meteoric materials including chondrule fragments. This may suggest
that some comets are contaminated by other populations. 

In the case of \textit{meteorites}, roughly the old differentiated
asteroids may be the population-IV origin, and the new non-differentiated
asteroids may be the population-VI origin. This differentiation is
ordinarily explained by the heat released from the radioactive decay
of $^{26}Al$ which was in the original asteroids \citet{Kita(2005),Luu(2015)}.
To distinguish the theories, we may need to analyze the origin of
Chondrite meteorites which include CAI and AOA chondrules. 

In the case of \textit{asteroids}, many observational results suggest
that the Asteroid belt is quite contaminated\citet{Ivezi=000107(2001),Mainzer(2011),DeMeo(2013)}.
This contamination in the asteroid belt by primordial Trans-Neptunian
objects was ordinarily explained by the Nice models and other variant
theories \citet{Crida(2009)}. On the other hand, according to our
model, the contamination is primordial, and therefore Planet migration
nor any hypothetical wind which mixes the primordial disk are not
needed, although quantitative analysis would be needed. 

\subsection{\label{subsec:Our-Earth--}Our Earth }

In the case of Earth, it is clear that the population is II in our
model, the same as the other rocky planets: Mercury, Venus, and Mars.
They are all shot from the 0.04AU orbit whose equilibrium temperature
is about 1400K, as shown above. Since most of the rocks melt at 800-1200K,
all these planets begun from the magma-ocean state. Some of them are
shot only once, but some others may shot multiple times which yields
extra contamination in composition. Therefore, the initial element
composition may slightly vary from planet to planet depending on the
number of slingshots. 

From the observations, Earth was formed $4.55\left(\pm0.01\right)\times10^{9}$
years ago\citet{Manhesa(1980)} and then, Moon was made $4.51\left(\pm0.01\right)\times10^{9}$
years ago\citet{Barboni(2017)}. During these initial periods, it
is proposed that Earth was in a magma-ocean state caused by Giant
impacts \citet{Zahnle(2007),Tucker(2014)}. Later Earth monotonically
cools down without any reheating such as Late Heavy Bombardment \citet{Cavosie(2019),Mojzsis(2019)}. 

Another recent observation \citet{Schiller(2020)} reports that the
Earth has a bare fraction of $\mu^{54}Fe$ $(<1ppm)$ compared with
most of the chondrites $(5-30ppm)$ except Cl chondrite$(<1ppm)$.
This fact conflicts with the popular idea that the Earth is made by
the stochastic collision and the accumulation of many kinds of rocks,
which have various fractions of $\mu^{54}Fe$. On the other hand in
our model, Earth is instantly formed by the rapid dust accumulation
and the slingshot in the time scale argued in subsection \ref{subsec:rocky-planets--}.
If the dust and dust clusters at OEIV ($r=0.04AU$) were dominated
by $\mu^{54}Fe$ poor Cl-like objects, then this observation may be
consistent with our model. 

\section{Conclusions}

We started our study by defining the planets directly based on the
laws of physics and the fundamental interactions. We found that the
non-metal ball defines Jupiter and the metal ball defines Neptune.
Planets are the stable structure bounded by gravity against the quantum
pressure and the atoms are by electromagnetism against the same. This
similarity has promoted us to seek for the populations of planets
as well as in the case of atoms. Thus we could start our present research. 

We started our study from the adoption of present many problems in
the planet formation theories based on the consideration that the
planets are universal and therefore actually forms without any intricate
artificial theories. Thus, instead of trying to challenge some of
these problems, we try to find any common trend in those problems.
We found that the dust and planets are quite dynamical objects that
fall and move in the protoplanetary disk. 

Thus we constructed our model assuming a single assumption that the
central gas void prevents them to fall into the central star. The
outer edge of this inner void (OEIV) turns out to be an unstable place.
Hot-Jupiters (HJ) are easily formed in the dense and trained dust
environment at OEIV and slingshot other bodies outward. Multiply falling
objects are scatted again outward by HJ. In this context, HJ on OEIV
are sheepdogs keeping watch over the falling objects and shooting
them back outward. 

Although we were initially afraid that the model would be quite exotic,
it turned out that all the processes in the model, in section 3, are
individually well-studies processes in the literature. Thus our contribution
in this paper would be systematic recombination of them. 

We summarize the further speculations of our model in order, although
the detail and the verification of them would need further study. 

\textbullet{} Dust aggregates in the protoplanetary disk fall toward
the OEIV where the aggregate density is increased and their motion
is coherently trained by frequent mutual collisions. Further, the
equilibrium temperature there is about 1400K and the aggregates stick
with each other to\textit{ grow to mm-cm size} easily. 

\textbullet{} These cm-sized bodies further grow to larger bodies
in the coherent, hot, and dense environment at OEIV. Their \textit{initial
chemical composition} is fixed at this stage. 

\textbullet{} Gas-giant planets form if heavier rocky clusters are
slingshot outward while the gas of the disk remains. Their core composition
would be almost the same as Earth, and the outside is simply the accumulated
gas and water which shows many phases depending on their temperature
at the radius. 

\textbullet{} Earth is a typical rocky planet shot from 0.04AU and
therefore water and other volatiles were not inherent at all. They
are later acquired material by the asteroid/comet collisions. The
chemical composition of the original Earth just formed is almost the
same as the cores of other Gas-giants. 

\textbullet{} Element composition of Earth would be non-uniformly
altered by giant impacts. However, in the context of Moon formation,
the giant impacts would not be necessary: Earth was originally hot
lava when shot from 0.04AU, and this gravitational shot might add
some extra angular momentum that caused the fission of Earth to separate
Moon. 

\textbullet{} The final composition of the atmosphere of a planet
would be a complicated mixture of volatile gas spouting from the original
planet and the acquired gas secondary controlled by the gravitational
stripping. 

\textbullet{} Super-Earth planets are typical metal-ball systems formed
by the balance of the quantum pressure and the electromagnetic force
of metals (in the seance of astronomy). This belongs to the pop.II
or III. 

\textbullet{} Even a binary star system can have a similar planetary
system as we have discussed provided that the binary orbit of them
is sufficiently larger than the size of the inner void regions. In
this case, we expect both stars have their OEIV which repel the approaching
dust or clusters outward. If the binary orbit is equal to or smaller
than the size of inner void regions, then the OEIV becomes strongly
unstable and may not work as a sheepdog. There would be no planetary
system in this case. 

We hope we can report the study on the detail and the verification
of the above soon. 

\acknowledgements{We wish to acknowledge Makiko Nagasawa at Kurume University, Tetsuro
Taki, Aya Higuchi, Takahiro Ueda, and Eiichiro Kokubo at NAOJ, Hideaki
Mouri at Meteorological Research Institute and in particular Shigeru
Ida at ELSI for useful discussions, valuable suggestions and encouragements.
We also thank the continuous support of the astrophysics group at
Ochanomizu University.}

\end{document}